# AI based Content Creation and Product Recommendation Applications in E-commerce : An Ethical overview

Aditi M Jain, Ayush Jain

Independent Researcher, USA

**A R T I C L E I N F O**

**A B S T R A C T**

As e-commerce rapidly integrates artificial intelligence (AI) for content creation and product recommendations, these technologies offer significant benefits in personalization and efficiency. AI driven systems automate product descriptions, generate dynamic advertisements, and deliver tailored recommendations based on consumer behavior, as seen in major platforms like Amazon and Shopify. However, the widespread use of AI in e-commerce raises crucial ethical challenges, particularly around data privacy, algorithmic bias, and consumer autonomy. Bias—whether cultural, gender based, or socioeconomic—can be inadvertently embedded in AI models, leading to inequitable product recommendations and reinforcing harmful stereotypes.

This paper examines the ethical implications of AI driven content creation and product recommendations, emphasizing the need for frameworks to ensure fairness, transparency, and need for more established and robust ethical standards. We propose actionable best practices to remove bias and ensure inclusivity, such as conducting regular audits of algorithms, diversifying training data, and incorporating fairness metrics into AI models. Additionally, we discuss frameworks for ethical conformance that focus on safeguarding consumer data privacy, promoting transparency in decision making processes, and enhancing consumer autonomy. By addressing these issues, we provide guidelines for responsibly utilizing AI in e-commerce applications for content creation and product recommendations, ensuring that these technologies are both effective and ethically sound.

## I. INTRODUCTION

Artificial intelligence (AI) is advancing at an accelerated pace, with many industries developing and adopting AI based solutions for a multitude of use cases. From healthcare to finance and retail, AI technologies are revolutionizing operational efficiencies and customer engagement. However, despite the rapid advancements, current AI systems often require manual human intervention to achieve optimal usability. This reality highlights the ongoing journey toward full automation, where the ethical challenge of ensuring an unbiased workflow independent of training data becomes increasingly critical.

E-commerce platforms are at the forefront of this AI revolution, employing technologies such as large language models (LLMs) and generative AI for various applications, including content creation, chatbots for personalized product recommendations, and ad-based marketing strategies. These technologies are pivotal for businesses seeking to enhance customer experiences by enabling seamless navigation through a vast array of products, services, and brands. However, as the reliance on AI increases, it is essential to consider the ethical implications of its deployment, particularly regarding bias in AI driven processes.

## II. E-COMMERCE AI MODELS AND THEIR TRAINING APPROACHES

We gather data from various applications of e-commerce sectors that utilize AI models in the year 2023 as presented in the heatmap in figure 1.

From this data, we then perform an analysis of the datasets used by these models, namely the following:

- Salesforce Einstein AI: Salesforce Einstein AI [1] primarily uses data from its CRM system, including customer interactions, sales data, and marketing insights. It relies heavily on historical customer engagement data to train its models for predictive analytics, sentiment analysis, and automation.
- Zendesk AI: Zendesk AI [2] is trained on customer service data, primarily supporting tickets, chat logs, and knowledge base articles. The AI uses this data for text classification, sentiment analysis, and chatbot training to enhance customer support experiences.

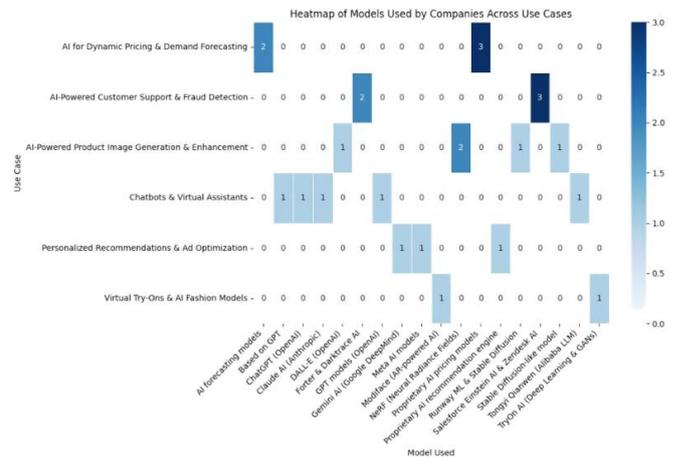

Fig. 1. Heatmap of AI models in E-commerce sectors (2023)

- Forter AI: Forter's AI [3] is trained primarily on e-commerce transaction data, including both legitimate and fraudulent transactions, customer behavior patterns, and historical fraud events. It uses machine learning techniques like supervised learning, anomaly detection, and reinforcement learning to detect fraud in real time.
- Darktrace AI: Darktrace's AI [4] is trained using network traffic data, endpoint logs, system behavior data, and historical cybersecurity threats. It uses unsupervised learning to detect anomalies and identify potential cybersecurity threats, including malware, ransomware, and insider attacks.
- Claude Models (Anthropic): Anthropic's Claude models (Claude 1, 2, and 3) [5] rely on a mixture of publicly available and licensed data. Their datasets include public domain books, scientific papers, code repositories, and high-quality web content, all filtered to maintain quality and reduce biases. However, these datasets are not immune to biases, especially when using large scale web data. As a result, the models can internalize patterns that may not accurately reflect the diversity of the global user base.

## III. HOW BIAS IS INTRODUCED IN E-COMMERCE APPLICATIONS

AI models in e-commerce rely on vast datasets sourced from consumer interactions, web data, and third-party providers. Bias emerges at multiple stages, including:

- Data Collection Bias: If training data is skewed towards a specific demographic, AI models may produce content and recommendations that favor certain groups while neglecting others.
- Algorithmic Bias: Machine learning models trained on imbalanced datasets inherit existing biases, leading to prejudiced product recommendations or misleading marketing content.
- Feedback Loop Bias: AI systems prioritize popular content and recommendations, reinforcing pre-existing consumer behavior and limiting diversity in suggested products.

E-commerce platforms heavily rely on specific AI models, like GPT, Claude, etc., for content creation, customer support, and personalized recommendations. These models are trained on vast amounts of data, though the exact specifics of their training datasets often remain unclear. Based on available information, these models are primarily trained on a mixture of publicly available data, licensed datasets, and proprietary content, which may include web data, academic papers, books, code repositories, and other sources. These models are also fine-tuned further with more specific training data to create a more specialized use case by various e-commerce companies. Hence, it becomes critically important to ensure that the training data has no bias.

## IV. HOW BIAS IS INTRODUCED THROUGH WEB-CREATED CONTENT

Web-created content, though abundant, can introduce significant biases in AI models, especially when the data is sourced from the internet. The issue arises from the inherent biases present in the content generated by humans. Graph presented in figure 2 provides a distribution of internet use cases by region. These are inherently inferred by these models, introducing bias regarding what kind of content attracts more customers for a specific product category.

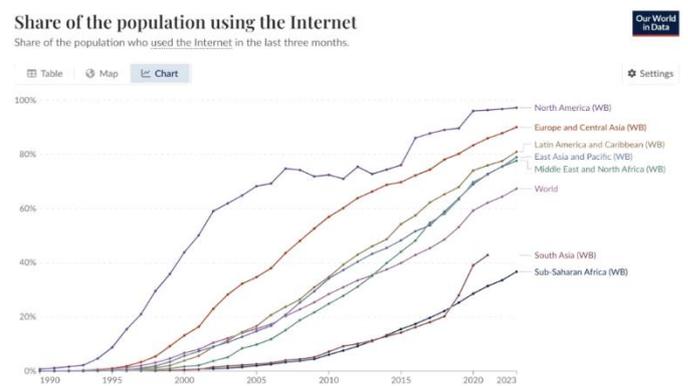

Fig. 2. Internet users by Demography

For instance, if an AI model is trained on a dataset dominated by web content from regions with higher internet penetration, like North America, it may learn cultural or behavioral patterns that do not represent the global population fairly. Take the example of internet usage data. In 2023, internet users from North America outnumber those from Sub-Saharan Africa by 2.5 times, as shown by the data from Our World in Data. This difference in internet usage patterns can skew the AI's understanding of various cultures as a result of this data collection bias.

For instance, if North American culture values reading books more heavily than Sub-Saharan African culture, an AI model trained predominantly on data from North America might infer that people from Sub-Saharan Africa have less interest in books. Over time, this could result in biased content suggestions, such as recommending fewer books to users from Sub-Saharan Africa, reinforcing the cultural imbalance. This ultimately brings forward algorithmic bias during fine-tuning and introduces a feedback loop bias over time.

## V. ENSURING FAIRNESS IN TRAINING DATA

To prevent biases, it is essential that models are not trained on "all" available data indiscriminately. Rather, training datasets should be carefully curated to ensure balance and diversity, minimizing the risk of perpetuating existing biases. A few key considerations when selecting training datasets for e-commerce models include:

A. Balanced Datasets

The dataset used to train models should reflect a wide range of cultural, regional, and behavioral differences to prevent the model from internalizing skewed perspectives. For instance, the inclusion of data from Sub-Saharan Africa, Asia, and other underrepresented regions is crucial for preventing bias.

B. Maintaining Exploration in Model Training

AI models should have the ability to explore new avenues and suggestions beyond the most prevalent patterns seen in training data. This helps ensure that the AI does not become overly reliant on the majority behavior seen during training but continues to offer diverse and unbiased recommendations.

C. Human Oversight

While AI models can process large datasets, human oversight is indispensable to identify and correct biases that may emerge during training. Having humans in the loop ensures that content generation and model behavior remain in line with fairness and ethical standards.

## VI. ETHICAL FRAMEWORKS

In recent years, the e-commerce sector has experienced unprecedented growth, driven by advancements in technology and changing consumer behaviors. As businesses increasingly turn to AI to enhance operational efficiency, improve customer engagement, and deliver personalized experiences, the ethical implications of these technologies have come to the forefront. The integration of AI in e-commerce not only offers the potential for significant competitive advantages but also raises critical questions regarding transparency, accountability, and fairness. As companies harness AI algorithms to analyze consumer data and automate decision making processes, the risk of bias and discrimination becomes a pressing concern. Moreover, the collection and utilization of vast amounts of personal data necessitate a careful consideration of privacy rights and ethical responsibilities. To navigate these challenges, it is essential for e-commerce businesses to adopt comprehensive ethical frameworks that prioritize consumer trust and societal values. Such frameworks will help ensure that AI technologies are deployed responsibly and equitably, fostering a sustainable digital marketplace that benefits both businesses and consumers. Here are some relevant ethical frameworks tailored to these areas:

A. Transparency and Honesty

E-commerce platforms must prioritize transparency and honesty in their content and advertising practices to build consumer trust. This means providing clear, accurate, and honest information about products and services, including any potential limitations or risks associated with them. Transparent advertising also involves disclosing when content is sponsored or when influencers are promoting a product in exchange for compensation. By doing so, platforms ensure that consumers make well informed decisions. Content creators and advertisers have a responsibility to avoid exaggerations or misleading claims that could distort consumer perceptions or lead to poor purchasing decisions. For instance, product descriptions should be truthful and comprehensive, reflecting the actual features and limitations of the product or service. Research shows that consumers' perception of value is significantly influenced by clear and truthful communication, as it fosters trust and strengthens customer relationships [6].

B. Informed Consent

Informed consent is a fundamental aspect of ethical data collection in e-commerce platforms, particularly regarding the use of consumer data for content creation and targeted advertising. Users should be fully informed about how their data will be collected, stored, and utilized, and platforms must obtain explicit consent before gathering or using this data. This transparency is crucial for building trust with users and ensuring that their privacy is respected. E-commerce platforms should provide clear and accessible privacy policies that explain the scope and purpose of data collection. Consent mechanisms, such as opt-in forms, should be used to ensure that users are aware of their rights and the extent to which their personal data will be used, especially in personalized advertising. Ethical concerns surrounding data privacy are growing, and businesses must take measures to ensure that they uphold users' rights in the digital age [7].

C. User Privacy and Data Protection

User privacy is a cornerstone of ethical e-commerce, and companies must take robust steps to protect consumer data from unauthorized access and misuse. E-commerce platforms are entrusted with sensitive personal information, such as payment details, addresses, and browsing behaviors, and have a responsibility to ensure that this data is used responsibly and securely. Strong data protection measures, such as encryption, secure storage, and access controls, must be implemented to safeguard users' privacy. Additionally, businesses should regularly conduct audits and comply with data protection regulations, such as the General Data Protection Regulation (GDPR), to ensure ongoing compliance and transparency. By adhering to privacy and data protection laws, companies not only avoid legal consequences but also foster consumer confidence in their platforms [8].

D. Fairness and Non-Discrimination

E-commerce platforms must strive for fairness and non-discrimination in their algorithms for ad targeting and content recommendations. Biases in algorithms can result in certain groups being unfairly targeted or excluded, potentially reinforcing harmful stereotypes or perpetuating discrimination. It is essential to ensure that algorithms do not inadvertently promote biased or discriminatory content, as this can have detrimental effects on users and harm the reputation of the platform. Regularly reviewing algorithms for fairness, equity, and transparency is critical, as is ensuring that all users, regardless of their demographic background, are treated equally. By implementing strategies that encourage diversity in marketing practices and promoting inclusive content, businesses can contribute to a more ethical and inclusive digital ecosystem [9].

E. Accountability and Responsibility

E-commerce platforms bear a significant responsibility for the content they promote and the ads they display. It is essential for platforms to establish clear accountability mechanisms to ensure that harmful or misleading content is identified and appropriately addressed. This includes regularly monitoring content for violations of ethical standards, such as deceptive advertising or manipulative tactics. Platforms should have procedures in place for taking corrective actions, including removing false claims, issuing retractions, and penalizing those who violate advertising guidelines. Clear accountability frameworks help foster trust with users and ensure that advertisers and content creators adhere to ethical guidelines.

Companies must remain vigilant in their oversight of third-party content to ensure that their platforms do not inadvertently spread harmful or misleading messages [10].

F. Inclusivity and Accessibility

Inclusivity and accessibility are critical aspects of ethical e-commerce design. E-commerce platforms should ensure that their applications are accessible to all users, regardless of their physical abilities, language, or cultural background. This involves making content, websites, and advertisements usable for individuals with disabilities, as well as creating interfaces that are culturally sensitive and easy to navigate for people from diverse backgrounds. Accessible design considerations include providing text alternatives for images, offering content in multiple languages, and ensuring that websites are navigable using screen readers or other assistive technologies. By adhering to established guidelines, such as the Web Content Accessibility Guidelines (WCAG), platforms can ensure that their content is available to the widest possible audience, contributing to a more equitable and inclusive digital environment [11].

G. Consumer Empowerment

Consumer empowerment is about providing users with the tools and information necessary to make informed decisions about their purchases and their online experience. In e-commerce, this includes offering clear disclosures about how ads are targeted and what personal data is collected. Consumers should have control over their data, including the ability to opt-out of personalized advertising or access their data upon request. By providing these options, platforms enhance consumer autonomy and trust. Giving users control over their information allows them to make decisions that align with their preferences and values, thus promoting transparency and ethical business practices. Empowering consumers in this way fosters long term relationships based on trust and mutual respect [12].

H. Ethical Marketing Practices

Ethical marketing practices are essential for e-commerce platforms seeking to build long term, trust based relationships with consumers. These practices should prioritize the well being of consumers over short-term profit maximization. This includes avoiding manipulative tactics that pressure consumers into making impulsive or ill-advised decisions. Instead, platforms should focus on promoting products and services in a way that respects consumer autonomy and values, emphasizing honest communication and customer satisfaction. Ethical marketing practices also involve respecting consumer preferences, fostering transparency in advertising, and providing clear and truthful product information. By adhering to ethical marketing standards, e-commerce platforms can build consumer loyalty and strengthen their reputation in the marketplace [13].

## VII. BEST PRACTICES FOR ETHICAL AI USE IN E-COMMERCE

After deriving inspiration from these ethical frameworks, we now outline best practices for ethical AI use in e-commerce. Implementing these practices is essential for ensuring that AI technologies contribute positively to the consumer experience while safeguarding their rights and fostering trust.

A. Implement Data Protection Measures

To ensure compliance with data privacy regulations such as the GDPR, CCPA, and others, businesses must implement robust data protection measures. This includes employing encryption, anonymization, and strict access controls to safeguard consumer data. Prioritizing data protection not only helps companies comply with legal standards but also fosters trust among consumers, which is essential for long term business success [14]. Regular audits and continuous monitoring of data management practices can also identify vulnerabilities, ensuring data security remains top-notch.

B. Conduct Regular Bias Audits

AI systems should be evaluated regularly for biases that could lead to unfair or discriminatory outcomes in content creation and advertising. Conducting bias audits helps identify patterns where certain groups may be unfairly targeted or excluded. Companies should take corrective action by retraining algorithms and incorporating diverse datasets to ensure that AI systems operate equitably. Research has shown that bias in AI systems can lead to significant societal harms, underscoring the importance of addressing this issue [15].

We can reduce bias by following best practices like:

- Diverse Data Sources: Ensure that training data is collected from a wide range of sources, representing various demographics, cultures, and viewpoints. This helps reduce the risk of over-representing any single group.
- Data Curation: Carefully curate training datasets to remove or minimize biased content. This can involve identifying and filtering out harmful stereotypes, hate speech, or misinformation.
- Bias Audits: Conduct regular audits of training data to identify and quantify biases. Use statistical analysis and bias detection tools to assess how different demographic groups are represented.
- Synthetic Data Generation: Create synthetic data to balance representation. This can involve generating text that reflects underrepresented perspectives or cultures without perpetuating stereotypes.
- Fine-tuning with Care: When fine-tuning models, use carefully selected datasets that focus on reducing bias. This can involve using datasets that are explicitly designed to promote fairness and inclusivity.
- Bias Mitigation Techniques: Implement algorithmic bias mitigation techniques during training. For example, adversarial debiasing can be used to adjust model predictions based on identified biases.
- Feedback Loops: Establish mechanisms for continuous feedback from users and stakeholders. This can help identify biases that emerge during deployment and inform future training efforts.
- Interdisciplinary Collaboration: Involve experts from various fields, including social sciences, ethics, and community representatives, to provide insights on bias and fairness.
- Transparency and Documentation: Maintain clear documentation of data sources, selection criteria, and any known biases in the datasets. Transparency helps users understand potential limitations and biases in the model.
- User Education: Educate users about the limitations of LLMs, including potential biases, and encourage critical thinking when interpreting model outputs.

C. Enhance Transparency

To build consumer trust, organizations must clearly communicate the role of AI in content creation and advertising. Transparency involves informing users about how AI algorithms influence the content they see, from personalized ads to news articles. By providing clear, accessible explanations of AI systems' decision making processes, companies can foster an environment of trust and reduce consumer skepticism. Studies have demonstrated that consumers are more likely to engage with platforms that offer insight into their algorithmic processes [16].

D. Respect Consumer Autonomy

Consumers should not be manipulated by marketing tactics that exploit their vulnerabilities. Ethical advertising practices emphasize respect for consumer autonomy by providing clear, accurate information that enables informed decision making. This involves avoiding misleading ads and respecting consumer preferences for privacy. By prioritizing autonomy, companies can build loyalty and positive brand reputation, which are essential in an era of heightened consumer awareness [17].

E. Establish Accountability Frameworks

AI driven decisions in advertising must be accompanied by clear accountability structures to ensure that companies take responsibility for their actions. This includes establishing protocols for tracking decision making processes and addressing any negative consequences resulting from AI-based advertisements. Companies should also designate accountability officers and create transparent reporting systems. Clear accountability frameworks promote ethical AI usage and minimize the risks of unintended harm [18].

F. Invest in Workforce Development

With the rise of AI and automation, there is a growing need for workforce development to prevent job displacement. Businesses should invest in reskilling and upskilling initiatives, helping employees transition to new roles that AI and automation are less likely to impact. This can involve offering training programs in emerging fields such as data science, AI ethics, and digital literacy. Research indicates that such initiatives not only support employees but also enhance organizational resilience in a rapidly changing job market [19].

G. Adopt Sustainable Practices

Incorporating sustainability considerations into AI deployment strategies is crucial for minimizing the environmental impact of technology. AI systems require significant computational power, which can lead to high energy consumption. By adopting energy-efficient technologies, using renewable energy sources, and optimizing algorithms for sustainability, companies can reduce their carbon footprint. Sustainable AI practices are not only ethically important but can also improve a company's reputation among environmentally-conscious consumers [20].

VIII. CONCLUSION

In conclusion, the ethical landscape of e-commerce content creation and product recommendations is complex, particularly regarding the introduction of bias and its implications for consumer trust and fair marketing practices. Bias can manifest in various forms, including algorithmic bias in ad targeting and content recommendations, which often leads to the reinforcement of stereotypes or the exclusion of certain demographic groups. This not only undermines the integrity of e-commerce practices but also diminishes consumer confidence in brands.

To mitigate bias, it is essential to adopt best practices that emphasize transparency, fairness, and inclusivity in content and advertising strategies. By implementing rigorous testing and auditing processes for algorithms, companies can identify and address biases that may affect their marketing efforts. Additionally, fostering an organizational culture that prioritizes diversity and inclusion in both content creation teams and target audiences can further reduce the risk of bias.

Several ethical frameworks, such as the principles outlined in the Belmont Report and guidelines from the American

Psychological Association, provide a solid foundation for establishing ethical standards in e-commerce. These frameworks encourage researchers and practitioners to prioritize informed consent, user privacy, and accountability, ensuring that consumer rights are respected throughout the content creation and advertising processes. Moreover, frameworks focused on fairness and user empowerment can guide e-commerce companies in developing practices that promote ethical marketing and responsible data usage.

Future studies will play a crucial role in enhancing these frameworks and adapting them to the rapidly evolving e-commerce landscape. By investigating the effectiveness of various ethical practices and developing new guidelines tailored to emerging technologies, researchers can contribute to a more ethical approach to content creation and ads optimization. As the digital marketplace continues to grow, the commitment to ethical considerations will be paramount in fostering consumer trust and ensuring that e-commerce serves as a fair and inclusive space for all.